\documentstyle[12pt,epsf]{article}

\def\fun#1#2{\lower3.6pt\vbox{\baselineskip0pt\lineskip.9pt
\ialign{$\mathsurround=0pt#1\hfil##\hfil$\crcr#2\crcr\sim\crcr}}}

\newcommand{\be}{\begin{eqnarray}}
\newcommand{\ee}{\end{eqnarray}}
\newcommand{\bd}{\begin{displaymath}}
\newcommand{\ed}{\end{displaymath}}
\newcommand{\ba}{\begin{array}}
\newcommand{\ea}{\end{array}}
\newcommand{\bt}{\begin{tabular}}
\newcommand{\et}{\end{tabular}}

\input epsf

\begin{document}

\begin{flushright}
ITEP-PH-3/98
\end{flushright}

\begin{center}
\large
{\bf CHIRAL PERTUBATION THEORY WITH LATTICE REGULARIZATION.}
\end{center}

\begin{center}
{\bf I.A. Shushpanov} 
\vspace{0.3cm}

and 
\vspace{0.3cm}

{\bf A.V. Smilga} 
\vspace{0.5cm}

\it{Institute for Theoretical and Experimental Physics,
B. Cheremushkinskaya 25, Moscow 117218, Russia}

\end{center}

\vspace{1.cm}
\centerline{June 1998}
\vspace{1.5cm}

\begin{abstract} 
We study the structure and perform some explicit calculations
for the power ultraviolet divergences and their renormalization
in the Chiral Pertubation Theory with lattice regularization 
in 1 and 2 loops.
\end{abstract}

\section{ Introduction.}

Chiral Pertubation Theory (ChPT) is nowadays a standard tool to 
study low-energy hadron dynamics $\cite{GL}$. The lagrangian
of ChPT involves
only lightest pseudoscalar degrees of freedom. The hadron excitations
with higher mass appear only indirectly, via phenomenological
constants of ChPT. As is well known, QCD enjoys an approximate
 $SU_L (N_f) \otimes SU_R (N_f)$ flavor chiral symmetry. This symmetry
is exact in the limit of massless quarks. When quark masses are not 
zero, this symmetry is broken, but, while the masses are small, this 
breaking is under control and can be taken into account by a systematic
expansion in mass. The lagrangian of ChPT is written in terms of the 
unitary matrix $U$ which is usually parameterized as 
$U=\exp \{2i\phi^a t^a/F_\pi \}$ ($\phi^a$ are the pseudoscalar meson
fields and $F_\pi$ is the pion decay coupling constant) and involves
all terms which are invariant under the chiral symmetry 
$SU_L (N_f) \otimes SU_R (N_f)$  and the non-invariant terms including
quark mass matrix ${\cal M}$ and its higher powers. The theory is not
renormalizable and involves an infinite number of counterterms which 
can be ordered by powers of derivatives (i.e. characteristic momenta)
and quark mass. To the leading order 
\be
\label{LCPT} 
L^{(2)} =\frac{F^2}{4} {\rm Tr} \{ \partial_\mu U^+ \partial_\mu U \}\ 
+ \ \Sigma {\rm Re Tr} \{{\cal M} U^\dagger \} \ ,
\ee 
where $\Sigma=|<\bar q q>_{\rm vac}|$ is the quark condensate.
The lagrangian ($\ref{LCPT}$) involves a dimensional coupling
$F \equiv F_\pi$ resulting in power ultraviolet divergences in 
the loops. The only way to get rid of them is to take into account
counterterms with higher derivatives and higher powers of
${\cal M}$ (this is what non-renormalizability
means). To perform the renormalization in practice, we have first 
to regularize the divergent integrals in ultraviolet. A naive momentum
cut-off does not work here; it breaks chiral invariance explicitly.
An almost exclusive way to handle the theory used so far in the literature
is the dimensional regularization. It is practically quite convenient
but has a drawback of being artificial and unphysical. In particular,
it does not display power ultraviolet divergences (which are there in 
any physical scheme) whatsoever. 

In this paper, we explore the lattice regularization scheme. As the only
way to define what quantum theory actually means is to put it on the 
lattice (so that path integrals involve a finite number of variables),
this regularization is the most natural and the most physical one from the 
"philosophical" viewpoint. In contrast to the dimensional regularization,
it does involve power ultraviolet divergences
explicitly which we consider as an
advantage rather than a drawback. The price to be paid is the absence of
Lorentz invariance in the regularized theory which makes the calculations
somewhat more tricky than with dimensional regularization. But this
technical complication is not really so serious. 

Intermittently,
the issue of lattice regularization and the related issue of the effects
due to nontrivial measure in the path integral for the chiral theory was 
addressed in the literature $\cite{HasBiet}$,
but we are not aware of any explicit
calculation with this scheme. We will perform here some calculations
with lattice regularization in 1 and 2 loops. For simplicity, we will
assume ${\cal M}$ to be diagonal ${\cal M} = m\hat 1$ and restrict
ourselves to the case $N_f=2$.

\section{ One-loop calculations.}

Lattice version of the Euclidean action, corresponding to 
the chiral lagrangian (\ref{LCPT}) is following:
\be
\label{LLCPT}
S=\frac{F^2 a^2}{2} \sum_{n,\mu} {\rm Tr} \{1-U^+_{n+e_\mu} U_n \}-
m\Sigma a^4 \sum_n {\rm Tr} \{U_n \} \ ,
\ee
where $a$ is the lattice spacing and $e_\mu$ are the unit four--vectors
connecting the adjacent nodes [writing the action (\ref{LLCPT}),
 we used the fact 
that Tr$\{ U\}$ is real when $U\in SU(2)$]. 
 Many different parameterizations of the matrix U can be chosen. 
To begin with, we choose the Weinberg parametrization:
\be
\label{U}
U=U_\mu \tau_\mu \quad \quad [\tau_\mu =(1,i\tau^a ); \quad
U^2_\mu=1; \quad U^a=\pi^a /F] \ .
\ee
We may expand then lattice action ($\ref{LLCPT}$) in the pion fields
$\pi^a$. For 1-loop calculations, it suffices to restrict oneself
by the terms $\sim \pi^2 \equiv \pi^a \pi^a$ and $\sim \pi^4$:
\be
\label{PICPT}
S=\sum_{n,\mu} \left[ \frac{a^2}{2}(\pi^a_{n+e_{\mu}} - \pi^a_n)^2 +
\frac{a^2}{4 F^2} (\pi^4_n -\pi^2_{n+e_{\mu}} \pi^2_n)\right] +
\frac{M^2_0 a^4} {2} \sum_n \left(\pi^2_n +\frac{1}{4 F^2} \pi^4_n \right) \ ,
\ee
where $M^2_0=2m\Sigma /F^2$  is the tree--level pion mass.

To the order $1/F^2$ there is only one relevant graph depicted
in Fig. 1

\begin{figure}
\centerline{\epsfbox{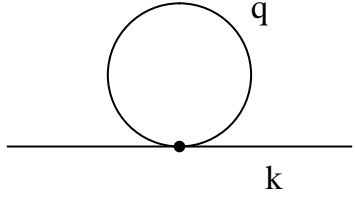}}
\caption{One--loop graph for $\Pi(k)$.}
\end{figure}
To calculate it, we need to know the free scalar propagator
on the lattice which can be taken e.g. from Ref.$\cite{Creutz}$ :
\be
\label{frprop}
D(x-y)=\int^{\pi /a}_{-\pi /a} \frac{d^4 q}{{(2\pi)}^4} \frac
{e^{iq(x-y)}}{M^2_0+\frac{2}{a^2}\sum_{\mu} [1-\cos(q_\mu a)]} \ ,
\ee
where $x_\mu=n_\mu a$, $y_\mu = n'_\mu a$ are the positions of lattice 
nodes and the momentum $q_\mu$ is restricted to lie in the interval
$|q_\mu| \le \pi /a$.

Calculating the polarization operator in Fig. 1, we obtain
 \be
 \label{Pik}
 \Pi^{\rm Fig.1}(k) \ =\ 
  \frac 2{a^2F^2} \left\{ 
  \sum_\mu [D(0) - \cos(k_\mu a) D(a e_\mu)] + \frac 54 M_0^2 a^2 D(0)
  \right\}
 \ee
 where
\be
\label{D(0)}
D(0)=\int^{\pi /a}_{-\pi /a} \frac{d^4 q}{(2\pi)^4} 
\frac{1}{M_0^2 + \frac{2}{a^2} \sum_\nu(1-\cos(q_\nu a))}\ ,
\ee

\be
\label{D(a)}
D(a e_\mu )=\int^{\pi /a}_{-\pi /a} \frac{d^4 q}{(2\pi)^4} 
\frac{\cos (q_\mu a)}{M_0^2 + \frac{2}{a^2} \sum_\nu(1-\cos(q_\nu a))} \ .
\ee
Here $D(a e_\mu )$ is the propagator connecting adjacent 
lattice nodes. It is the same in all directions. When $a \ll M_0^{-1}$,
 \be
 \label{Das}
 D(0) \ \sim \ \frac I{a^2} \  , \nonumber \\
 D(ae_\mu) \ \sim \ \frac 1{a^2} \left( I - \frac 18 \right) 
 \ee
where
\be
I=\frac{1}{2} \int^{\pi}_{\pi}
\frac{d^4 x}{(2\pi)^4} \frac{1}{\sum_i (1-\cos x_i)} =
\frac{1}{2}\int^\infty_0 ds e^{-4s} I^4_0 (s)
\approx 0.154934
\ee
[$I_0 (s)=1/2\pi \int^\pi_{-\pi} e^{s\cos x} dx$ is a Bessel
function].

Eq.(\ref{Pik}) is not, however, the final result for the polarization 
operator. Actually, it would be a disaster if it would.
In the chiral limit $M_0 = 0$, $\Pi^{\rm Fig.1}(k = 0) = 
 1/(a^4 F^2) \neq 0$. 
That would mean that the massless on the tree level pions 
acquire a mass (involving a quartic ultraviolet divergence)
after renormalization. 
That would break chiral invariance [manifest
in the lattice action (\ref{LLCPT})] which looks like a nonsense.
 The paradox is resolved by taking into account
the fact that the path integral in chiral theory contains the
integration over $SU(2)$ matrices $U$ with the invariant Haar measure
(otherwise the 	quantum theory would not be chirally invariant,
indeed). We have:
\be
\label{mes}
\int [dU]=\int \Pi_{n,a} \frac{dU^a_n}{U^0_n} \sim \int \Pi_{n,a}
\frac{dU^a_n}{1-\frac{1}{2} U^2_n} \sim \int \Pi_{n,a} d\pi^a_n
\exp \left\{- \sum_n \ln \left(1-\frac{1}{2}\pi^2_n /F^2\right)\right\} \ .
\ee
Expanding logarithm, we obtain the extra term $- \frac 1{2F^2} \sum_n \pi_n^2$
in the effective lattice action and, correspondingly,
the following  contribution in $\Pi(k)$
\be
\label{massmes}		
\Delta \Pi(k) [{\rm measure}]\ = \ -\frac{1}{a^4 F^2}
\ee
[in the continuum limit, the contribution from the measure has
the form $\sim F^{-2} \delta^{(4)} (0)$].
We see that the quartic divergent additive contributions to the pion
mass coming from the 1-loop graph in Fig.1 and those 
from the measure cancel out completely.

The quadratic ultraviolet divergences in $\Pi(k)$ survive, however. Adding
the contributions (\ref{Pik}) and (\ref{massmes}) and retaining the leading
terms, we obtain
 \be
 \label{Piexp}
\Pi(k) \ =\   \frac 1{a^2F^2} \left[
\frac {3IM_0^2}2 + k^2\left( I - \frac 18 \right)  
  \right] \ + \ o\left( \frac{1}{a^2} \right) \ .
 \ee

We can now go over into the Minkowski space (with $k_E^2 \to -k_M^2,
\ \ \Pi_E \to -\Pi_M$)
and find the renormalized value of the mass as the solution to
the dispersive equation
$k_M^2 = M_0^2 + \Pi_M(k_M^2)$. We obtain
  \be
\label{M}
M^2_{phys}=M_0^2 \left[1+\frac{1}{2 F^2a^2 }\left(I + \frac{1}{4}\right) 
 + \cdots \right]\ ,
\ee
The residue of the propagator at the pole is
\be
\label{Z}
Z = 1  - \frac{1}{F^2 a^2} \left(   I - \frac{1}{8}\right) + 
\cdots \ .
\ee

Let us do the same with an arbitrary parameterization.
Define new fields $\phi^a$ so that
\be
\label{param}
\pi^a=\phi^a +C \phi^a \frac{\phi^2}{F^2}+\cdots \ .
\ee
Note that the choice $C=-1/6$ corresponds to the exponential 
parametrization $U=\exp \{i\tau^a \phi^a /F \}$.

When substituting ($\ref{param}$) in ($\ref{PICPT}$), the quadratic
terms are not changed and the terms $\sim \phi^4$ have the form:
\be
\label{PHICPT}
S^{(4)} = \frac{a^2}{F^2}\sum_{n,\mu} \left[\left(2C+\frac{1}{4}\right)
\phi^4_n - \frac 14 \phi^2_{n+e_{\mu}} \phi^2_n 
- 2C\phi^2_n \phi^a_n \phi^a_{n+e_{\mu}} \right] +
\frac{M^2_0a^4}{8 F^2}\sum_n (1+8C)\phi^4_n \ .
\ee

In a general parameterization, the contribution from the loop graph in 
the pion polarizational operator is
\be
\Pi^{\rm Fig.1}(k) = \frac{2}{a^2F^2} 
\left\{ 
  \sum_\mu [(1 + 20C)D(0) - \cos(k_\mu a) D(a e_\mu)
  - 10C D(ae_\mu) -  \right.  \nonumber \\ 
  10C \cos(k_\mu a) D(0)] + 
\left.  \frac 54 (1+ 8C) M_0^2 a^2 D(0) \right\} \ = \nonumber \\
  \frac{1}{a^4 F^2}(1+10C) + \frac 1{a^2F^2} 
  \left\{ IM_0^2 \left(\frac 32 + 10C \right) +
  k^2 \left[ I(1+ 10C) - \frac 18 \right] \right\}
  + \  o\left(\frac{1}{a^2} \right) \ .
\label{pol1op}
\ee
The contribution from the measure is
\be
\label{massmesprparam}
\Delta M^2_0 [{\rm measure}]=-\frac{1}{a^4 F^2}(1+10C) \ ,
\ee
i.e quartic divergences coming from the measure and additive divergent
contribution from the graph in Fig.1 cancel  each over in any 
parameterization.
\footnote{A particular choice $C=-1/10$ where no quartic divergences
in mass appear whatsoever is especially convenient.}

Solving again the dispersion equation  with the polarization operator 
($\ref{pol1op}$),  one can readily see that the pole position of 
the propagator does not depend on the parameterization parameter C
 to the order $\sim 1/F^2$ and is given by the relation (\ref{M}).
Nothing prevents, however, for the residue of the pion propagator to depend
 on the parameterization, and it does:
\be
\label{ZC}
Z=1 - \frac{1}{F^2 a^2} \left[(1+10C)I - \frac{1}{8}\right] 
+ \cdots \ .
\ee

Let us calculate now the renormalization of the effective chiral 
coupling constant $F$. The latter is defined as the residue
of the correlator of flavor non-singlet axial currents $A^a_\mu (x)$ 
at the pion pole 
\be
\int d^4 x e^{ikx} <A^a_\mu (x) A^b_\nu (x) > \sim F^2 
\frac{k_\mu k_\nu}{k^2 -M^2_\pi} \delta^{ab} \ .
\ee

Technically, it is somewhat more convenient to pick up the 
coefficient of the structure $\sim g_{\mu\nu}$
in the correlator; due to the chiral current conservation
in the chiral limit, the coefficients of these two tensor
structures are the same.

There is only one relevant 1-loop diagram depicted in Fig.2.

\begin{figure}
\centerline{\epsfbox{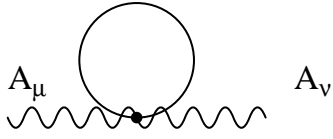}}
\caption{Renormalization of $F_\pi$.}
\end{figure}


It involves the vertex $<A_\mu A_\nu |\pi\pi>$which can be
found by "covariantizing" the derivatives in Eq.($\ref{LCPT}$) 
by the rule
$\partial_\mu U \rightarrow \partial_\mu U -i(A_\mu U + U A_\mu)$
with $A_\mu \equiv A_\mu^a t^a$.
\footnote{This is all very much parallel to conventional
calculations in chiral theory with dimensional regularization
$\cite{GL}$
or when calculating the effects due to 
finite temperature $\cite{temp}$
(see recent $\cite{temp1}$
for review) or to an external magnetic field $\cite{SS}$.}
We have $\Delta F^2 = - 2D(0)$ and hence, in the chiral limit,
  \be
  \label{F}
  F^2_{phys} \ = \ F^2_0 \left[1 -  \frac{2I}{F^2_0 a^2}+\cdots \right]
   \ee 
in any parameterization.

 \section{Two-loop calculations.}

We will calculate here only the additive two-loop corrections
$\sim 1/F^4 a^6$ in the pion mass in the chiral limit 
($M^2_0 \equiv 0$) and show that, after taking into account
carefully all the contributions (including the relevant contributions
from the measure), the final result is zero. We will work in the
Weinberg parameterization.

To do the calculation, we are in a position to expand the action
($\ref{LLCPT}$) up to the terms $\pi^6 /F^4$ which have the form:
\be
\label{L6}
S^{(6)} \ =\  \frac{a^2}{8 F^4} \sum_{n,\mu} 
(\pi^6_n -\pi^4_n \pi^2_{n+e_\mu})\ .
\ee

\begin{figure}
\centerline{\epsfbox{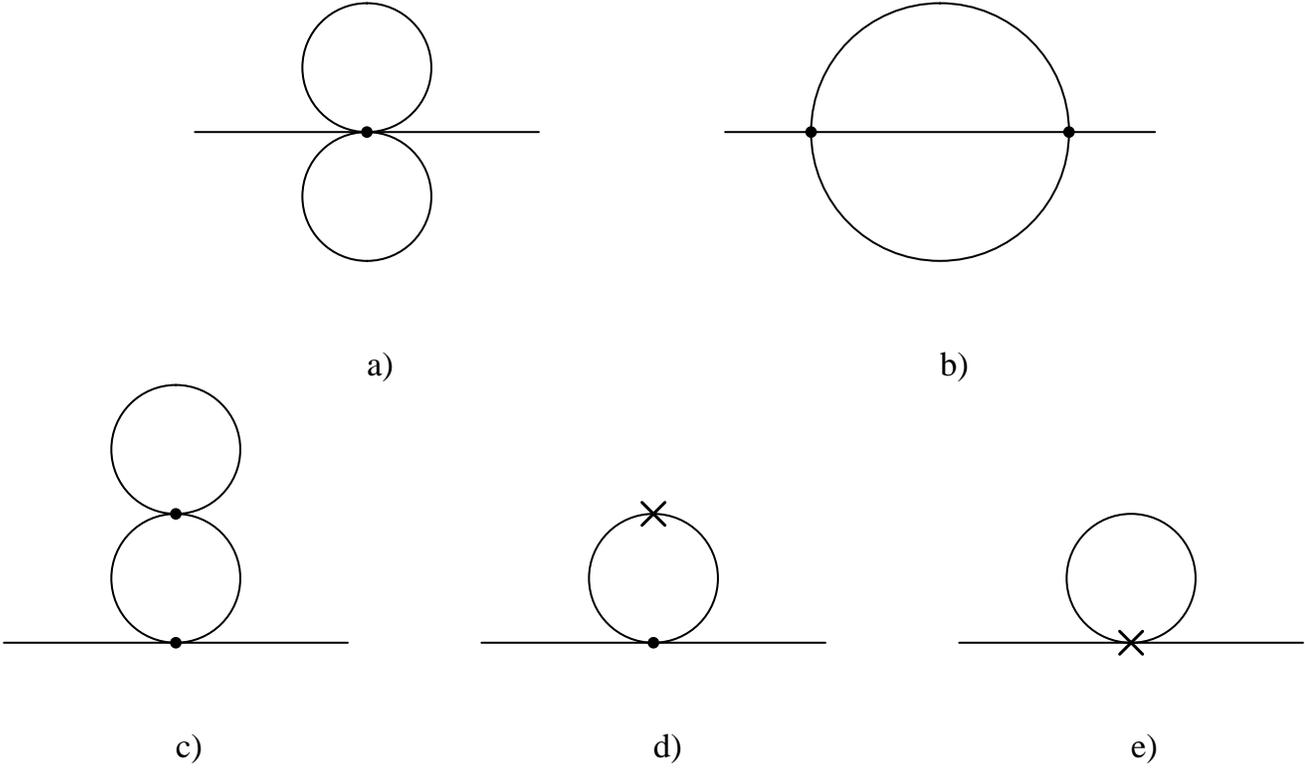}}
\caption{Polarization operator in two loops.}
\end{figure}


The relevant two loop graphs are depicted in Fig.3 where crosses
stand for the measure factor. Extracting the 6-pion vertex from
Eq.($\ref{L6}$) and substituting the free lattice propagator
($\ref{frprop}$), the contribution of the graph in Fig.3a in 
the pion polarization operator is:
\be
\label{Pa}
\Pi^a (k^2)=\frac{5}{a^2 F^4} \sum_\mu [3 D^2(0)- D^2 (a e_\mu)-
2\cos(k_\mu a) D(0)D(a e_\mu )] \ .
\ee
Setting here $k = 0$ and substituting the propagators (\ref{D(0)}, 
\ref{D(a)}) in the form (\ref{Das}), we find the
additive divergent contribution
in the mass:
\be
\label{ma}
\Delta ({M_\pi^2})^{Fig.3a}=\frac{10}{a^6 F^4} \left(I-\frac{1}{32}\right)\ .
\ee 
  
The graph in Fig.3b can be calculated using the 4-pion vertices from 
Eq.($\ref{PICPT}$). 
Putting $k=0$ from the very beginning, we obtain

\be
\label{Pb}
\Pi^{\rm Fig.3b} (0)= \nonumber \\
-\frac{1}{4 a^6 F^4} \sum_{\mu,\nu} 
\int^{\pi}_{-\pi} \frac{d^4 x d^4 y}{(2\pi)^8}
\frac{5-10\cos (x+y)_\mu  +2\cos x_\mu  \cos y_\nu +
3\cos(x+y)_\mu  \cos (x+y)_\nu }
{ \sum_{\alpha,\beta,\gamma} (1-\cos x_\alpha )
(1-\cos y_\beta ) [1-\cos (x+y)_\gamma ]}
\ee
which can be rewritten as
\be
\label{Pbm}
\Pi^{\rm Fig.3b} (0) =-\frac{1}{a^6 F^4}\left(4I-\frac{3}{16}\right) -
\frac{1}{2 a^6 F^4} J 
\ee
where
\be
\label{zero}
J=\int^{\pi /a}_{-\pi /a} 
\frac{d^4 x d^4 y}{(2\pi)^8} 
\left[\frac{1}{\sum_\mu (1-\cos x_\mu)} -
\frac{1}{\sum_\mu (1-\cos (x+y)_\mu)}\right]
\left[1-\frac{2}{\sum_\nu (1-\cos y_\nu)} \right] \ .
\ee
Integrating over $d^4 x$ first and using the fact that the integration
region covers the whole period of the cosine functions (so that 
the shift $x_\mu \rightarrow x_\mu + y_\mu$ does not change the value
of the integral), we see that $J=0$. The corresponding additive
mass correction is given by the first term in ($\ref{Pbm}$):
\be
\label{mb}
(\Delta M_\pi^2)^{\rm Fig.3b}=-\frac{1}{a^6 F^4} \left(4I-\frac{3}{16} 
\right) \ .
\ee

The graphs in Fig.3c,d have the same structure as that of the 
1-loop graph in Fig.1, only we have to substitute there the 1-loop 
correction to the scalar propagator rather than the tree expression.
We obtain:
\be
\label{Pcd}
\Pi^{Fig.3c,d} (k^2)=\frac{2}{a^2F^2} \sum_\mu
[\Delta D(0)- \Delta D(a e_\mu ) \cos k_\mu a]
\ee
where
\be
\Delta D(x)=-\int\frac{d^4 q}{(2 \pi^4)} 
\frac{e^{iqx} \ \Pi^{(1)} (q)}{[\frac{2}{a^2}\sum_\mu 
(1-\cos q_\mu a)]^2}\ ,
\ee
with $\Pi^{(1)} (q)$ being given by the sum of the contributions
(\ref{Pik}) and (\ref{massmes}) at $M_0^2=0$ : 
$$ \Pi^{(1)}(q) \ =\ \frac {2D(ae_\mu)}{a^2F^2} \sum_\mu [1 - \cos(q_\mu a)]
\ = \ \frac {2(I - 1/8)}{a^4F^2} \sum_\mu [1 - \cos(q_\mu a)]\ . $$
We obtain
\be
\Delta D(0)\ =\ -\frac{1}{a^4 F^2}I \left(I - \frac{1}{8}\right), \nonumber \\
\Delta D(a e_\mu)\ =\ -\frac{1}{a^4 F^2}\left(I - \frac{1}{8}\right)^2\ .
\ee
Therefore
 \be
 \label{mcd}
 (\Delta M_\pi^2)^{\rm Fig.3c,d} = 
\Pi^{\rm Fig.3c,d}(0)  = -\frac{1}{a^6 F^4} \left(I - \frac{1}{8}\right) \ .
 \ee

Finally, the Fig. 3,e corresponds to the expansion of
the factor $1/U^0$
in the group measure ($\ref{mes}$) up to the terms $\sim \pi^4$
which effectively produce the following correction to action:
\be
S^{(6)}[{\rm measure}]\ =\ -\frac{1}{4 F^4} \sum_n \pi^4_n\ ,
\ee
and, correspondingly, to the 4-pion vertex. The calculation gives
\be
\label{me}
(\Delta M^2_\pi)^{\rm Fig.3e}=\Pi^{\rm Fig.3e} (0)=-\frac{5I}{a^6 F^4} \ .
\ee

Adding up the additive ultraviolet divergent contributions ($\ref{ma}$),      
($\ref{mb}$), ($\ref{mcd}$) and ($\ref{me}$) together,      
we find that total sum is zero, so that a massless pion 
stays massless, indeed.

Let us comment on the other possible contributions in the pion
polarization operator to the order $\sim 1/F^4$. They
come from the 1-loop graph in Fig.1 where the 4-pion vertex
is extracted not from the leading order lagrangian ($\ref{LCPT}$)
but from the higher derivatives terms:
\be
\label{L4}
L^{(4)} \ =\ \frac{1}{4} l_1 [{\rm Tr} \{\partial_\mu U^+ \partial_\mu U\}]^2
+\frac{1}{4} l_2 [{\rm Tr} \{\partial_\mu U^+ \partial_\nu U\}]^2
+\cdots 
\ee
where $l_{1,2}$ are some phenomenological parameters
(or, better to say, some infinite constants in the free
chiral lagrangian which will be transformed into finite 
phenomenological parameters after renormalization).
However, the 4-pion vertices following from Eq.($\ref{L4}$)
are quartic in momenta. One can be convinced that the vertex 
is zero for zero value of the external momentum and 
the corresponding contribution is proportional to the square 
of external momentum and affects only Z-factor,
not the additive term in the mass renormalization.

\section{Conclusions}

The method of calculations with the lattice regularization
developed here presents an alternative to the standard
dimensional regularization method. It is very explicit and 
physical and, though the calculations on the 4-dimensional
lattice are slightly more involved than the calculations
in the continuous space of the dimension $4-\epsilon$
(or, rather just less habitual), these complications are
not so serious. The main distinction compared to the 
dimensional regularization method is that power ultraviolet
divergences are not "swept under the carpet", but are
seen explicitly. Thereby, the main realm of application of
our method could be the problems where the structure of power
ultraviolet divergences in a non-renormalizable theory
needs to be analyzed.

We calculated explicitly quadratically divergent pieces
in $F_\pi$, $M_\pi$ and in the residue of the propagator Z
[see Eqs. ($\ref{F}$), ($\ref{M}$) and  ($\ref{Z}$)]. We did not
study the logarithmically divergent pieces whose structure
should be roughly the same as with dimensional 
regularization. It would be an interesting methodic problem
to verify it explicitly.

We are indebted to H. Leutwyler for illuminating discussions.
This work  was done under the partial support of 
RFBR INTAS grants 
93-0283, 94-2851, RFFI grant 97-02-16131, CRDF grant RP2-132,
and Schweizerischer National Fonds grant 7SUPJ048716.

\end{document}